\documentclass{PoS}

\newcommand{\ds}{\displaystyle}
\newcommand{\ident}{{\bf 1}}

\title{Axial and pseudoscalar form-factors of the $\Delta^+(1232)$}

\ShortTitle{$\Delta^+(1232)$ form-factors}

\author{Constantia Alexandrou\\
        Department of Physics, University of Cyprus, P.O. Box 20357,
        1678 Nicosia, Cyprus, and\\
Computation-based Science \& Technology Research Center, The Cyprus Institute, P.O. Box 27456, 1645 Nicosia, Cyprus\\
        E-mail: \email{alexand@ucy.ac.cy}}
\author{\speaker{Eric B. Gregory}\\
        Department of Physics, University of Cyprus, P.O. Box 20357,
        1678 Nicosia, Cyprus\\
        E-mail: \email{gregory.eric@ucy.ac.cy}}
\author{Tomasz Korzec\\
Institut f\"ur Physik,
   Humboldt Universit\"at zu Berlin, Newtonstrasse 15, 12489 Berlin,
   Germany\\
        E-mail: \email{korzec@physik.hu-berlin.de}}
\author{Giannis Koutsou\\
       J\"ulich Supercomputing Center, Forschungszentrum J\"ulich, D-52425 
       J\"ulich, Germany, and \\
       Bergische Universit\"at Wuppertal, Gaussstr. 20, D-42119 Wuppertal, 
       Germany\\
        E-mail: \email{i.koutsou@fz-juelich.de}}
\author{John Negele\\
        Center for Theoretical Physics, Laboratory for Nuclear Science and 
        Department of Physics, Massachusetts Institute of
        Technology, Cambridge, Massachusetts 02139, U.S.A.\\
        E-mail: \email{negele@mitlns.mit.edu}}
\author{Toru Sato\\
        Department of Physics, Osaka University, Osaka 560-0043, Japan\\
        E-mail: \email{tsato@phys.sci.osaka-u.ac.jp}}
\author{Antonios Tsapalis\\
       Hellenic Naval Academy, Hatzikyriakou Ave., Pireaus 18539, Greece, and \\
       Department of Physics, National Technical University of
       Athens, Zografou Campus 15780, Athens, Greece\\
       E-mail: \email{tsapalis@snd.edu.gr}}

\abstract{We present first results on the axial and pseudoscalar $\Delta$ form factors. 
The analysis is carried out in the quenched approximation where
 statistical errors are small and the lattice set-up
can be investigated relatively quickly. We also present an analysis 
with  a hybrid action using staggered sea quarks and domain-wall 
valence fermions.}

\FullConference{The XXVIII International Symposium on Lattice Field Theory, Lattice2010\\
		June 14-19, 2010\\
		Villasimius, Italy}

\begin{document}
\section{Introduction}
A major focus of interest in hadronic physics is the quest to understand from
first principles the structure of mesons and baryons. In particular 
form factors yield information about the size and shape of the hadrons.

Much theoretical and experimental 
 work has gone into understanding the structure of 
nucleons and the $N\rightarrow \Delta$ transition. Lattice QCD
calculations of the nucleon and $N\rightarrow \Delta$ form factors (FFs)
have been carried out within the same lattice setup as the one used in this 
work~\cite{Alexandrou:2003ea,Alexandrou:2006ru,Alexandrou2007}.
Experimental information on  the FFs of $\Delta(1232)$ is scarce due
to its short
lifetime ($\sim 10^{-23}$ s)~\cite{Kotulla:2002cg,LopezCastro:2000cv}. 
However, in a finite-volume simulation with heavy pions, the 
$\Delta$ is stable and accessible to lattice techniques.
 A pioneering lattice study~\cite{Leinweber:1992hy} investigated the 
electromagnetic form-factors of the $\Delta$ in the quenched approximation. 
Recently,  a state-of-the-art lattice calculation of the
electromagnetic FFs of the $\Delta$  and
the associated  transverse charge densities in the
infinite momentum frame  has been carried out~\cite{Alexandrou:2009hs}.
In this report we 
extend the effort to the axial and pseudoscalar form factors of the 
$\Delta(1232)$ and present preliminary results. To our knowledge this
is the first time that these FFs have been computed.

Despite the difficulty of experimental confirmation of these results,  
they can yield an evaluation of the axial charge and the effective 
$\pi\Delta\Delta$ couplings, parameters which can be fed into chiral 
expansions to aid the chiral extrapolations of, for example, the nucleon 
axial charge. The axial Ward-Takahashi identity relates the axial FFs to 
the pseudoscalar FFs. As in the nucleon case, one can derive the generalized 
Goldberger-Treiman relations. In this work we derive these and check their 
validity.

\section{Lattice calculation}
This project closely follows the methods used for extracting 
$\Delta^+$ electromagnetic form factors as described comprehensively in 
Ref.~\cite{Alexandrou:2009hs}. We begin with the expression of the isovector 
axial vertex:
\begin{equation}
\label{ax_decomp_min}
\langle \Delta(p_f,s_f) | A^\mu| \Delta(p_i,s_i)\rangle = 
\overline{u}_\alpha(p_f,s_f)
\left[{\mathcal O}^{\mu {\rm A }}\right]^{ \alpha\beta} 
u_\beta(p_i,s_i),
\end{equation}
with
\begin{equation}
A^\mu(x)= \overline{\psi}(x)\gamma^\mu\gamma_5 \frac{\tau^3}{2}\psi(x)\nonumber.
\end{equation}

The right-hand side is an expression containing the most general 
decomposition of the axial vertex in terms of four form-factors, which we 
label $g_1$, $g_3$, $h_1$ and $h_3$:
\begin{equation}
\label{ax_operator}
{\mathcal O}^{\mu {\rm A }}=
-g^{\alpha\beta}
\left(g_1(q^2)\gamma^\mu\gamma^5 
    + g_3(q^2) \frac{q^\mu}{2M_\Delta}\gamma^5\right)
+\frac{\ds q^\alpha q^\beta}{\ds 4M_\Delta^2}
\left(h_1(q^2)\gamma^\mu\gamma^5 
   + h_3(q^2) \frac{q^\mu}{2M_\Delta}\gamma^5\right),
\label{axial-vector}
\end{equation}
and $u_\alpha$ is the Rarita-Schwinger spinor and $q=p_f-p_i$.

Similarly we can write the pseudoscalar vertex in terms of two form-factors,
$\tilde{g}$ and $\tilde{h}$:
\begin{equation}
\label{ps_decomp_min}
\langle \Delta(p_f,s_f) | P| \Delta(p_i,s_i)\rangle = 
\overline{u}_\alpha(p_f,s_f)
\left[{\mathcal O}^{ {\rm PS }}\right]^{ \alpha\beta} 
u_\beta(p_i,s_i)
\end{equation}
with
\begin{equation}
P(x)= \overline{\psi}(x)\gamma_5 \frac{\tau^3}{2}\psi(x)
\label{pseudo-scalar}
\end{equation}
and
\begin{equation}
{\mathcal O}^{{\rm PS}}=
-g^{\alpha\beta}
\tilde{g}(q^2)\gamma^5 
+\frac{\ds q^\alpha q^\beta}{\ds 4M_\Delta^2}
\tilde{h}(q^2)\gamma^5 
\end{equation}
We  isolate the form-factors by constructing 
ratios of lattice two- and three-point functions. 
The standard lattice interpolating field with the $\Delta^+$ 
quantum numbers is given by
\begin{equation}
   {\bf\chi}^{\Delta^+}_{\sigma\alpha}(x) = \frac{1}{\sqrt{3}} \epsilon^{abc}
\Bigl[2\left({\bf u}^{a\top}(x) C\gamma_\sigma {\bf d}^b(x)\right)
 {\bf u}_\alpha^c(x)  
+ \left({\bf u}^{a\top}(x) C\gamma_\sigma {\bf u}^b(x)\right) {\bf d}_\alpha^c(x
)\Bigr]\, .
\end{equation}
The two-point and three-point functions of interest are:
\begin{eqnarray}
G_{\sigma\mu\tau}^{X}(\Gamma^\nu,\vec{q},t)&=& 
\sum_{\vec{x},\vec{x}_f}e^{+i\vec{x}\cdot\vec{q}}
\Gamma^\nu_{\alpha^\prime\alpha}
\langle\chi_{\sigma\alpha}(t_f,\vec{x}_f)X_\mu(t,\vec{x})\overline{\chi}_{\tau\alpha^\prime}(0,\vec{0})\rangle\\
G_{\sigma\tau}(\Gamma^\nu,\vec{p},t)&=& 
\sum_{\vec{x}_f}e^{-i\vec{x}_f\cdot\vec{p}}
\Gamma^\nu_{\alpha^\prime\alpha}
\langle\chi_{\sigma\alpha}(t,\vec{x}_f)\overline{\chi}_{\tau\alpha^\prime}(0,\vec{0})\rangle
\end{eqnarray}
where $X$ stands for the axial-vector and pseudo-scalar currents defined 
in Eqs. (\ref{axial-vector}) and (\ref{pseudo-scalar}) respectively and 
\begin{equation}
   \Gamma^4 = \frac{1}{4}(\ident  + \gamma^4)\, , \qquad \qquad \Gamma^k =
   \frac{i}{4}(\ident  + \gamma^4)\gamma_5\gamma_k\, , \qquad k=1, 2, 3\, .
\end{equation}
We examine ratios of these to eliminate unknown $Z$-factors
and leading time-dependence.:
\begin{equation}
        R_{\sigma\mu\tau}^{X}(\Gamma,\vec{ q},t) 
= \frac{G_{\sigma\mu\tau}^{X}(\Gamma,\vec{ q},t)}
{G_{k k}(\Gamma^4,\vec 0, t_f)}\ \sqrt{\frac{G_{kk}(\Gamma^4,\vec p_i, 
t_f-t)G_{kk}(\Gamma^4,\vec 0  ,t)G_{kk}(\Gamma^4,\vec 0,t_f)}
                                                    {G_{kk}(\Gamma^4,\vec 0, 
t_f-t)G_{kk}(\Gamma^4,\vec p_i,t)G_{kk}(\Gamma^4,\vec p_i,t_f)}}\, 
\label{ax_ratio},\nonumber
\end{equation}
These ratios tend to a constant at large Euclidean time separations $t_f-t_i$ and $t$:
\begin{equation}
\label{generic_ratio}
R_{\sigma(\mu)\tau}(\Gamma,\vec{q},t)^X \stackrel{{\tiny\begin{array}{c}t_f-t\rightarrow\infty\\t_f-t_i\rightarrow\infty\end{array}}}{\longrightarrow} C\Pi_{\sigma(\mu)\tau}^X
=C{\rm tr} \left[\Gamma \Lambda_{\sigma\sigma^\prime}{\mathcal O}^X_{\sigma^\prime(\mu)
\tau^\prime}
\Lambda_{\tau^\prime\tau}\right],
\end{equation}
with the kinematical constant given by
\begin{equation}
C\equiv\sqrt{\frac{3}{2}}\left[\frac{2 E_{\Delta(p_i)}}{M_\Delta} 
                          +\frac{2 E^2_{\Delta(p_i)}}{M^2_\Delta} 
                          +\frac{  E^3_{\Delta(p_i)}}{M^3_\Delta} 
                          +\frac{  E^4_{\Delta(p_i)}}{M^4_\Delta} \right]^{-\frac{1}{2}}.
\end{equation} 
\section{Lattice simulation}
\begin{table}[h]
\small
\begin{center}
\begin{tabular}{ccccccc}
\hline\hline 
\multicolumn{7}{c}{Wilson fermions}\\
\hline
V& \# confs & $\kappa$ & $m_\pi$ & $m_\pi/m_\rho$ & $m_N$ & 
$m_\Delta$ \\ 
& & & (GeV)& & (GeV) & (GeV)\\ 
\hline
\multicolumn{7}{c}{SIM-I : Quenched, \quad $\beta=6.0,~~a^{-1}=2.14(6)$~GeV} \\
\hline
$32^3\times 64$& 200 & 0.1554 &0.563(4)& 0.645(9)  & 1.267(11) & 1.470(15)\\
$32^3\times 64$& 200 & 0.1558 &0.490(4)& 0.587(12) & 1.190(13) & 1.425(16)\\
$32^3\times 64$& 200 & 0.1562 &0.411(4)& 0.503(23) & 1.109(13) & 1.382(19)\\
\hline
\hline
\multicolumn{7}{c}{SIM-II: Mixed action} \\
\multicolumn{7}{c}{
Asqtad ($am_{\mbox{\tiny u,d/s}} = 0.01/0.05$),  
DWF ($ am_{\mbox{\tiny u,d}} = 0.0138$), $a^{-1} = 1.58(3)$~GeV}\\
\hline
$28^3\times 64$ &200 &  & 0.353(2) & 0.368(8)&1.191(19) &
 1.533(27)\\
\hline\hline
\end{tabular}
\end{center}
\caption{Ensembles and parameters used in the calculation of form factors.}
\label{Table:params}
\end{table}

We utilize 200 quenched Wilson configurations on a lattice of size
$32^3\time 64$ at
$\beta=6.0$, which corresponds to inverse lattice spacing of
  $a^{-1}=2.14(6)$ GeV. We perform the analysis at three hopping parameter 
values $\kappa=0.1554$, $0.1558$ and 
$0.1562$, corresponding to pion masses $m_\pi = 563, 490,$ and $411$ MeV respectively.
Additionally, we use a mixed action of domain wall valence fermions on 
a staggered sea simulated by the  MILC collaboration  
with an Asqtad improved action~\cite{Bernard:2001av}, with a pion mass of
$353$ MeV.
A total of 200 configurations are analyzed at one value of
the pion mass. The details of the simulations are summarized in 
Table \ref{Table:params}.

 We use the sequential 
source method~\cite{Dolgov:2002zm} to calculate three-point functions.
We use the same fixed source-sink separation as was used in Ref.~\cite{Alexandrou:2009hs} of $\sim 1$ fm, or 11 time-slices on the three quenched ensembles 
and eight time-slices on the mixed-action ensemble.

\section{Extracting form-factors}
For each ratio 
\begin{equation}
\Pi_{\sigma(\mu)\tau}^X
={\rm tr} \left[\Gamma \Lambda_{\sigma\sigma^\prime}{\mathcal O}^X_{\sigma^\prime(\mu)\tau^\prime}
\Lambda_{\tau^\prime\tau}\right],\nonumber
\end{equation}
we work out the trace algebraically for specific combinations of 
$\sigma$, $\tau$, and $\Gamma^j$ or $\Gamma^4$:
\begin{equation}
\Pi^{X,I}_\mu(q)= \sum_{j=1}^3\sum_{\sigma,\tau =1}^3 T_{\sigma\tau}{\rm tr}\left[\Gamma^j \Lambda_{\sigma\sigma^\prime}(p_f) {\mathcal O}_{\sigma^\prime \mu \tau^\prime}\Lambda_{\tau^\prime\tau}(p_i)\right]
\end{equation}
or 
\begin{equation}
\Pi^{X,II}_\mu(q)= \sum_{\sigma,\tau =1}^3 \tilde{T}_{\sigma\tau}{\rm tr}\left[\Gamma^4 \Lambda_{\sigma\sigma^\prime}(p_f) {\mathcal O}_{\sigma^\prime \mu \tau^\prime}\Lambda_{\tau^\prime\tau}(p_i)\right]
\end{equation}
with
\begin{equation}
T_{\sigma\tau}=\left[
\begin{array}{rrr}
1&0&0\\
0&1&0\\
0&0&1
\end{array}
\right]
\qquad
{\rm and} 
\qquad 
\tilde{T}_{\sigma\tau}=
\left[
\begin{array}{rrr}
0&1&-1\\
-1&0&1\\
1&-1&0
\end{array}
\right].
\end{equation}
We refer to these traces as Type I and Type II respectively.
For the right-hand sides we now have linear combinations of 
the form-factors where the coefficients are functions of $E_i$ the initial 
energy, $M_\Delta$ and the spatial initial momentum $p_i$. 
(We Wick rotate and work in the rest-frame of the sink). In general, 
the form of the expression is different for $\mu=4$ and $\mu=1,2,3$.
The left hand side is calculated on the lattice, so we may now solve a 
system of linear equations to isolate the form-factors. The FFs are extracted 
by the simultaneous over-constrained analysis of all the relevant ratios that 
contribute to the transition per given $Q^2$. 
The renormalization constant $Z_A$ is required for the axial FFs. They are
given in Table IV in \cite{Alexandrou2007}.

\begin{figure}
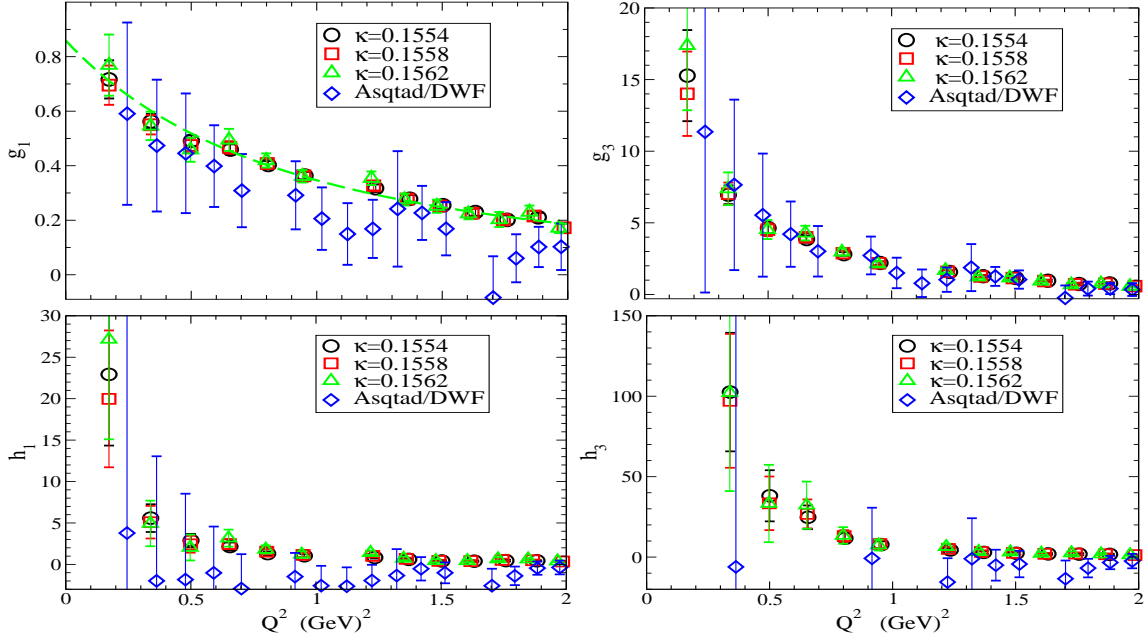

\begin{center}
\includegraphics[width=.497\textwidth,height=1.6in]{g1_ax_all.eps}
\hspace{0.025in}
\includegraphics[width=.485\textwidth,height=1.6in]{g3_ax_all.eps}
\vspace{-0.2in}
\includegraphics[width=.497\textwidth,height=1.7in]{h1_ax_all.eps}
\includegraphics[width=.497\textwidth,height=1.7in]{h3_ax_all.eps}
\end{center}
\caption{Results for the four axial form-factors, $g_1$, $g_3$, $h_1$ and 
$h_3$. The dashed curve shows consistency of $g_1$ for 
the $\kappa=0.1562$ ensemble with a dipole fit giving 
a pole mass of $1.67(5)$GeV. }
\label{axial_ff_figs}
\end{figure}

We summarize the results obtained for the axial form factors calculated on 
all four ensembles in 
Figure \ref{axial_ff_figs}. By extrapolating the $g_1$ curve to $Q^2=0$
we get an estimate of the axial charge of the $\Delta^+$. 
In order to connect $g_1(0)$ to the axial charge $g_{\Delta\Delta}$, we use the 
relations given in \cite{Jiang:2008we}. We find  
that $g_1(0)= -\frac{1}{3}g_{\Delta\Delta} $. Using the tree-level $SU(4)$ 
relation $g_{\Delta\Delta}=-\frac{9}{5}g_A$
and the experimental value $g_A=1.2694(28)$ \cite{PDG} we expect that 
\begin{equation}
g_1(0)\approx\frac{1}{3}\frac{9}{5}(1.27)= 0.76,
\end{equation}
which is consistent with our results.

\section{Effective couplings}
Referring to Eq.~(\ref{ps_decomp_min}), we decompose the $\Delta$ matrix 
elements of the pseudoscalar current into two axial
$\pi\Delta\Delta$ couplings, $G_{\pi\Delta\Delta}$ and $H_{\pi\Delta\Delta}$, with
the relation:
\begin{equation}
\label{eff_couple_eq}
2m_q\langle\Delta_{p_f}|P| \Delta_{p_i} \rangle  \equiv 
\left( \frac{m_\Delta^2}{E_\Delta(\vec{p}_f)E_\Delta(\vec{p}_i)}\right)
\frac{2f_\pi m_\pi^2 }{(m_\pi^2-q^2)}
\left[g^{\alpha\beta}G_{\pi\Delta\Delta}(q^2)
+\frac{q^\alpha q^\beta}{4m_\Delta^2}H_{\pi\Delta\Delta}(q^2)
\right]
\overline{u}_\alpha \gamma^5 u_\beta
\end{equation}
and we identify:\\
\begin{equation}
m_q\tilde{g}\equiv\frac{f_\pi m_\pi^2G_{\pi\Delta\Delta}(q^2)}{(m_\pi^2 - q^2)}
\qquad
{\rm
and
}
\qquad
m_q\tilde{h}\equiv\frac{f_\pi m_\pi^2H_{\pi\Delta\Delta}(q^2)}
{(m_\pi^2 - q^2)}
\end{equation} 
Note that because the two different possible contractions of the Dirac indices 
of the $3/2$-spinors give us two pseudoscalar form-factors  we get  
two effective axial couplings, unlike the case of the nucleon and the
$N-\Delta$ transition. The quark mass $m_q$ is  computed from the axial 
Ward-Takahashi identity and $f_\pi$ from the pion-to-vacuum amplitude. 
Both are
computed from appropriate combinations of two-point functions as shown in 
the reference \cite{Alexandrou2007}. The renormalization factor $Z_P$ is not 
required as its occurrences in $m_q$ and $\tilde{g}$ or $\tilde{h}$ cancel.

\begin{figure}
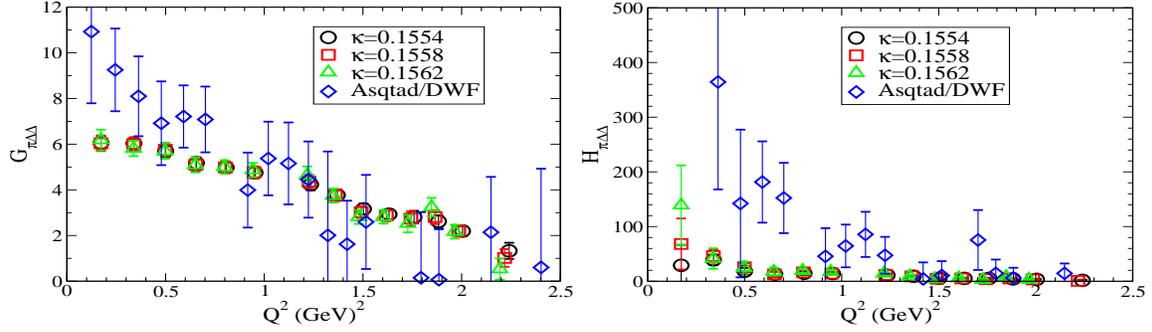

\begin{center}
\includegraphics[width=.497\textwidth,height=1.7in]{GpDD_all.eps}
\includegraphics[width=.497\textwidth,height=1.7in]{HpDD_all.eps}
\end{center}
\caption{The effective couplings $G_{\pi\Delta\Delta}$ and $H_{\pi\Delta\Delta}$
for the quenched ensemble with $\kappa=0.1554$.}
\label{eff_couplings}
\end{figure}

The results for these form-factors are shown in 
Fig.~\ref{eff_couplings}. As can be seen, despite the large statistical errors, $G_{\pi\Delta\Delta}$
 increases with decreasing $Q^2$ for the unquenched lattices.

\section{Goldberger-Treiman relations}
\begin{figure}
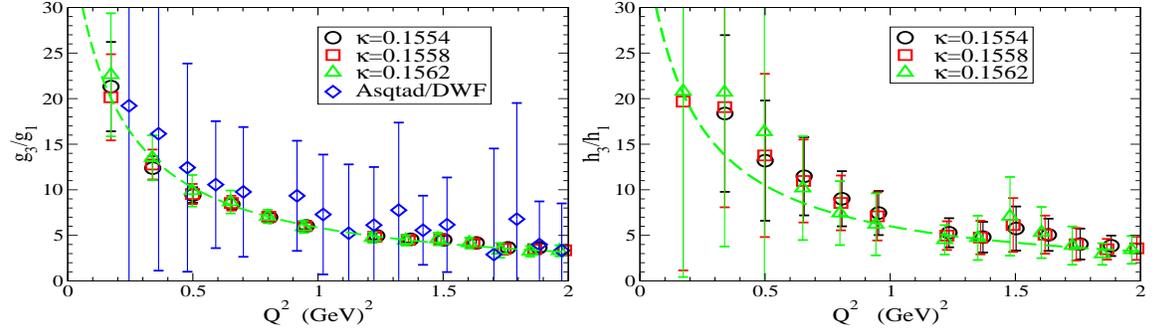

\begin{center}
\includegraphics[width=.497\textwidth,height=1.7in]{g3g1_rat_ax_all.eps}
\includegraphics[width=.497\textwidth,height=1.7in]{h3h1_rat_ax_all.eps}
\end{center}
\caption{Ratios of axial form-factors, $g_3/g_1$ and $h_3/h_1$. Dashed curves 
show consistency with a pion pole fit for the 
$\kappa=0.1562$ ensemble. We omit the mixed ensemble from the $h_3/h_1$ plot as
the signal is washed out by the larger error bars.}
\label{axial_ff_ratios_figs}
\end{figure}
From the axial Ward-Takahashi identity, we have the relationship
\begin{equation}
\langle\Delta| \partial_\mu A^\mu | \Delta \rangle = 2m_q\langle\Delta|P | \Delta
 \rangle.
\end{equation}
Applying the momentum operator on (\ref{ax_decomp_min}) and (\ref{ax_operator}),
the left-hand side gives 
\begin{equation}
\langle\Delta|\partial_\mu A^\mu | \Delta \rangle=
2m_\Delta
\left[
(g_1 - \tau g_3)g^{\alpha\beta} +
(h_1 - \tau h_3)\frac{q^\alpha q^\beta}{4m_\Delta^2}
\right]
\overline{u}_\alpha \gamma^5 u_\beta
\end{equation}
with $\tau=\frac{-q^2}{(2m_\Delta)^2}$. Using Eqs. (\ref{eff_couple_eq}) 
we get
\begin{equation}
2m_\Delta\left(g_1  -\tau g_3\right) =\frac{2f_\pi m_\pi^2 
G_{\pi\Delta\Delta}(q^2) }{(m_\pi^2-q^2)}
\qquad
{\rm and}
\qquad
2m_\Delta\left(h_1  -\tau h_3\right) =\frac{2f_\pi m_\pi^2 
H_{\pi\Delta\Delta}(q^2) }{(m_\pi^2-q^2)}.
\end{equation}

 If we demand that the $g_3$ and $h_3$ terms cancel the pole at $q^2=m_\pi^2$,
we get the Goldberger-Treiman relations. In Figure 
\ref{axial_ff_ratios_figs} we show that the ratios $g_3/g_1$ and
$h_3/h_1$ are consistent with  pion-pole behavior. We note that, as with the 
effective axial couplings, there are two Goldberger-Treiman relations, namely:
\begin{equation}
\label{GTR_g}
f_\pi G_{\pi\Delta\Delta}(q^2) = m_\Delta g_1(q^2)\nonumber
\qquad {\rm and} \qquad   
f_\pi H_{\pi\Delta\Delta}(q^2) = m_\Delta h_1(q^2)\nonumber.
\end{equation}
\begin{figure}
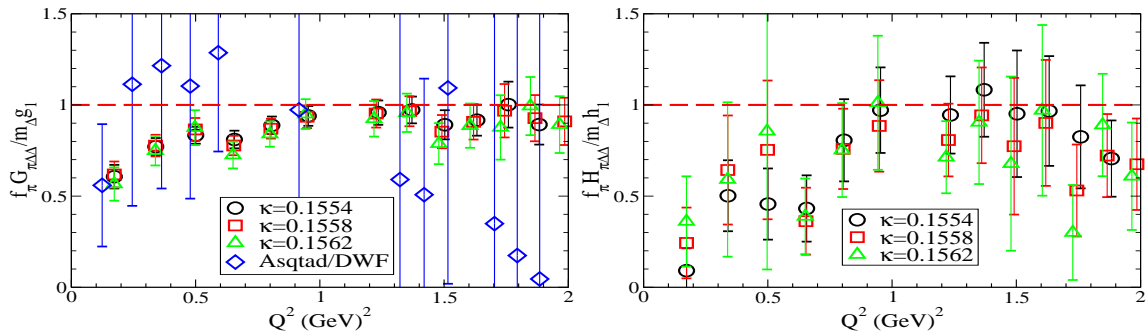

\begin{center}
\includegraphics[width=.497\textwidth,height=1.7in]{GTR_GpDD_all.eps}
\includegraphics[width=.497\textwidth,height=1.7in]{GTR_HpDD_all.eps}
\end{center}
\caption{The two Goldberger-Treiman relations for the $\Delta$ baryon. }
\label{GTR_figs}
\end{figure}

In Fig.~\ref{GTR_figs} we plot the ratio of the left-hand to right-hand 
sides of the expressions given in (\ref{GTR_g}). 
For low $Q^2$, these quenched ratios deviate from unity but are in agreement 
with unity for $Q^2\stackrel{>}{\sim}0.8 {\rm GeV}^2$. 
Similar behavior was observed for $G_{\pi NN}$ and $G_{\pi N\Delta}$ in 
\cite{Alexandrou2007}. In the unquenched ensemble the errors are too large 
to enable any conclusions.
\section{Conclusions}
In this work we have evaluated, for the first time, the $\Delta^+$ 
axial form factors, $g_1$, $g_3$, $h_1$, $h_3$ as well as
the pseudoscalar form factors $\tilde{g}$, $\tilde{h}$. 
 We have shown that these axial and pseudoscalar vertex 
compositions yield {\em two} effective $\pi\Delta\Delta$ couplings,
 $G_{\pi\Delta\Delta}$, and $H_{\pi\Delta\Delta}$, which in turn satisfy dual 
Goldberger-Treiman relations. Results obtained in the quenched theory are 
accurate enough to enable a check of these relations and show that there are 
deviations for small $Q^2$ values where chiral effects are expected to be 
large. Unquenched results using a mixed action have large statistical errors 
and require further analysis for allowing a definite conclusion to be reached.

\end{document}